# A predictive analytics approach to reducing 30-day avoidable readmissions among patients with heart failure, acute myocardial infarction, pneumonia, or COPD

Issac Shams, Saeede Ajorlou, Kai Yang

*Healthcare Systems Engineering Group, Wayne State University, Detroit, MI 48201*

**Abstract**

Hospital readmission has become a critical metric of quality and cost of healthcare. Medicare anticipates that nearly $17 billion is paid out on the 20% of patients who are readmitted within 30 days of discharge. Although several interventions such as transition care management have been practiced in recent years, the effectiveness and sustainability depends on how well they can identify patients at high risk of rehospitalization. Based on the literature, most current risk prediction models fail to reach an acceptable accuracy level; none of them considers patient's history of readmission and impacts of patient attribute changes over time; and they often do not discriminate between planned and unnecessary readmissions. Tackling such drawbacks, we develop a new readmission metric based on administrative data that can identify potentially avoidable readmissions from all other types of readmission. We further propose a tree-based classification method to estimate the predicted probability of readmission that can directly incorporate patient's history of readmission and risk factors changes over time. The proposed methods are validated with 2011–12 Veterans Health Administration data from inpatients hospitalized for heart failure, acute myocardial infarction, pneumonia, or chronic obstructive pulmonary disease in the State of Michigan. Results shows improved discrimination power compared to the literature (*c*-statistics > 80%) and good calibration.

# 1 Introduction

Hospital readmission is disruptive to patients and costly to healthcare systems. Unnecessary return to hospitals shortly after discharge has been increasingly perceived as a marker of the quality of care that patients receive during hospital admission [1]. About one in five Medicare fee-for-service beneficiaries, totaling over 2.3 million patients, are rehospitalized within 30 days after discharge, incurring an annual cost of $17 billion, which constitutes nearly 20% of Medicare's total payment [2]. However, it is reported by the Medicare Payment Advisory Commission (MedPAC) that about 75% of such readmissions can and should be avoided because they often result from a fragmented healthcare system that leaves discharged patients with preventable flaws such as hospital-acquired infections and other complications, poor

planning for follow up care transitions, inadequate communication of discharge instructions, and failure to reconcile and coordinate medications [3]. Variations in both medical and surgical readmission rates by different hospitals and different geographic regions indicate that some centers (or regions) perform better than others at containing readmission rates [2,4]. Studies also show that the adjusted readmission rate in the US is among the highest ranking in comparison to European countries [5].

In addition, effective October 2012, as directed by Patient Protection and Affordable Care Act (PPACA, also called Obamacare), the Centers for Medicare and Medicaid Services (CMS) started to cut reimbursement funds for hospitals that have excessive 30-day readmission rates for heart failure, acute myocardial infarction, or pneumonia patients. This included 2,213 US hospitals with approximately $280 million funds nationwide, which constitutes 1% of the total Medicare budget. Moreover, this cut will grow to 2% and 3% for FY 2014 and 2015, respectively, with four additional conditions such as chronic obstructive pulmonary disease and coronary bypass graft. As a result, numerous intervention programs have been proposed by policymakers and healthcare organizations to reduce rehospitalizations and improve quality and access to care [6].

While it would be perfect to include all patients in a transitional care intervention, due to their resource intensive nature on one hand and hospital supplies constraints on the other, it is inevitable to target and deliver such efforts to those subgroups that are at greater risk. Nevertheless, identifying patients at increased risk of readmission is challenging and calls for advanced analytics tools that help to stratify risk into clinically relevant classes and provide information early enough during the hospitalization. Various methods have been proposed in recent years to predict hospital readmission but most of them do not yield acceptable predictive accuracy, or they are based on patient factors that are not typically collected during clinical care [7]. Furthermore, a few methods have tried to distinguish avoidable readmission form all other types of readmissions [8], but it remains a disagreement how to systematically define and identify those readmissions that can be prevented based on credible clinical criteria.

In this paper, we propose a predictive analytics framework that enables medical decision makers to characterize and (more accurately) predict avoidable readmissions, and to investigate the effects of different patient risk factors on the likelihood of rehospitalization. The goal of our study is two-fold: (1) to develop and internally validate an administrative algorithm for characterizing avoidable readmissions from all types of readmissions, and (2) to create and validate a simple and real-time readmission risk prediction model that can produce more desirable prediction accuracy than the literature ($c$-statistics > 80%). We use, for model derivation and validation, all 2011–12 Veteran Affairs (VA) inpatient records after hospitalization for heart failure, acute myocardial infarction, pneumonia, or chronic obstructive pulmonary disease in four facilities of the State of Michigan.

# 2  Study design and methods

## 2.1  Data used

The dataset used in this retrospective cohort study is provided by the Veteran Health Administration (VHA), which is the largest single medical system in the United States, with 152 medical centers and nearly 1400 outpatient clinics. We analyze inpatient administrative records gathered from four medical facilities in the State of Michigan, namely, Ann Arbor, Battle Creek, Detroit, and Saginaw, to identify all hospitalizations for Heart Failure (HF), Acute Myocardial Infarction (AMI), Pneumonia (PN), and Chronic Obstructive Pulmonary Disease (COPD) from Fiscal Year 2011 to FY12. Cohorts are marked with ICD-9-CM (International Classification of Diseases, Ninth Revision, Clinical Modification) codes, similar to the coding utilized by the CMS for calculating hospital readmission rates. There were no major changes in the hospital bed supplies, and in the patient admission/discharge processes through that period of time. During a hospital stay, patients may move to different acute wards within the hospital and their episodes of care are carefully tracked with standard computerized means. We use additional data files for patients with chronic conditions as well as patients exposed to environmental hazards such as Agent Orange, to effectively illustrate those impacts on the risk of readmission.

The dataset set contains 7200 records that correspond to 2985 distinct adult patients with principal (or secondary) discharge diagnoses of HF, AMI, PN, and COPD (the original set includes 7237 records form which 37 are dropped since they have severe data quality issues). General exclusions include: (1) Hospital admissions within 24 hours of index discharge, (2) Hospitalizations with a length of stay less than 24 hours (observation stays) or followed by a death, (3) Patients transferred to another acute care facility, (4) Patients discharged against medical advice. To count readmissions in the last month of FY12, the first month of FY13 is taken into account. In additions, we omit stays in long term care, nursing home, psychiatry, rehabilitation, and hospice wards. However, as we are interested in modeling the effect of patient's related factor changes (over time) on the risk of readmission, unlike most studies in the literature [9-11], we do not exclude recurrent (re)admissions of the same patient from the analyses. We also design both internal and external model validations by using stratified split sample and bootstrap resampling methods.

*Controlled variables*

We aggregate patient level data files with provider and station levels in order to obtain various types of risk factors for this study. To achieve a better picture of the data environment, we further arrange them into five groups: (1) Demographics: age at discharge, sex, race, and marital status; (2) Socioeconomic:

means tested income, and insurance status (Medicare, Medicaid, private, none); (3) Utilization: length of stay of the index hospitalization (LOS), treating facility, source of admission (direct from home, outpatient clinic, transition from any of the four VA hospital, VA Nursing Home Care Unit (NHCU), and VA domiciliary), primary care provider, enrollment priority, and average distance (between patient's home zip code and the zip code of the facility he/she got admitted); (4) Service based: Agent Orange status, Prisoner Of War (POW) status, and radiation status; and (5) Comorbidity and severity: Diagnosis Related Group (DRG), Hierarchical Condition Category (HCC), and Care Assessment Need (CAN) score. The variables are selected based on the relevant medical literature and confirmed by a group of VA health professionals.

The enrollment priority is a priority level assigned according to the veteran's severity of service-connected disabilities and the VA means test. The DRG is a validated reimbursement classification scheme exploited to identify the cost of services that a hospital renders. In its basic version, the groups are organized with respect to their similarities in patient diagnosis, age, sex, and the presence of complications or comorbidities; then a measure of cost is attached to each group [12]. HCCs have been used *ad hoc*, mainly for case-mix and risk adjustment in healthcare utilization and payment systems. Each HCC group forms a set of clinically and cost-similar conditions reflecting hierarchies among related diseases as defined by the ICD-9-CM codes [13]. We create dummy variables for both the DRG and HCC variables in the regression studies; that is, if a patient is a member of the category, he or she is given a 1 on this variable; otherwise the score remains zero. The CAN score is a general illness severity score that reflects the likelihood of admission or death within a specified time period, and it works somewhat similar to diagnostic cost group (DxCG) risk score [14]. The score is commonly expressed as a percentile ranging from 0 (lowest risk) to 99 (highest risk) and it shows how a VA patient is compared with others pertaining to the chances of hospitalization or death. It is interesting to note that all predictor variables except length of stay are real time and would be available before patient discharge, so they can be employed in planning for pre-discharge (transitional care) intervention programs.

*Study outcomes*

The main outcome is 30-day avoidable readmission as defined later in Section 2.2; if no consecutive admission is occurred within 30 days after the most recent admission, the outcome is flagged as censored. Unlike large part of the literature that studies only the *occurrence* of readmission by logistic (or probit) regression methods [15,11], our current method is a hybrid of both *occurrence* and *timing* of readmission, which enables us to directly incorporate the effect of partially known inforamtion (censored observations) into the risk of readmission. We further modify the approach introduced by Goldfield et al. [16], to distinguish between avoidable and unpreventable outcomes. The most common causes of readmission for the four cohorts as well as their changes over time are also investigated as secondary health outcomes.

## 2.2 Methods

*Measuring potentially avoidable readmissions*

Measuring and counting readmissions have been questionable among health researchers, and a number of different metrics have been proposed that vary in some ways. For example, the time period encompassing recurrent admissions after discharge ranges from one week to 180 days, among which 30-day is the most common. Some metrics consider readmissions for any cause ("all-cause readmission") while others try to exclude subsequent admissions likely to be planned or unrelated to the initial admission [16]. Among the metrics, Centers for Medicare and Medicaid Services (CMS) 30-day readmission [17] and the 3M Corporation Potentially Preventable Readmissions [18] have been used more often. The main differences between them are related to the risk adjustment used and restriction to include only clinically related readmissions. More details about the rationale of these methods and their specific distinctions can be found in [19].

In this study, since our goal is more to develop and validate a risk prediction model that can be used for clinical applications (rather than hospital profiling and payment adjustment), we derive a hybrid approach adopting both the CMS and 3M rationales to choose from the patient outcomes. In a nutshell, we first apply the CMS method to exclude those planned procedures that are followed by a non-acute or a non-complication of care condition; then the 3M procedure is implemented on the remaining indices in order to extract potentially avoidable readmissions. However, we modify the exclusion criteria of both methods and implement VHA definitions of eligible discharge. To increase the overall precision of the proposal, we also got help from three reviewers to judge all cases identified, after completing each constituent algorithm. Moreover, instead of the APR DRG system, the newly-developed Diagnostic Cost Group Hierarchical Condition Category, Solution version, version 21 (DCG/HCC v21) is utilized to assess the clinical relationship between each readmission and its initial admission(s) [13]. We chose the DCG/HCC risk adjustment system because 1) it is a part of models that have been used and evolved over two decades of research; 2) it has special adjustments for elderly beneficiaries as well as patients with chronic conditions; and 3) it is recalibrated regularly according to recent modifications on diagnosis and expenditure data.

The algorithm, which we call Potentially Avoidable Readmission (PAR), is stated as follows:

**Step 1** (general inclusion/exclusion)

I. Identify heart failure (HF), acute myocardial infarction (AMI), pneumonia (PN), and COPD cohorts based on principal (or secondary) discharge diagnoses, and eliminate all other conditions. Merge records of the same patient if he/she had multiple hospitalizations on the same day to the same medical unit. This applies to both medical and surgical patients.
II. Establish 30-day readmission time interval and categorize each entry as either admission or readmission. Also, define *eligible admissions* as all admissions that are at risk of having a readmission.
III. Exclude:
 a) From the admission set, cases whose discharge status is "death," since they cannot have any readmission. These correspond to stand-alone admissions.
 b) From the admission set, cases whose discharge status are "transfer" to another acute care facility, except the four hospitals studied. The reason is that the hospital cannot affect a patient's consequent care under such circumstances. If transferred among the four hospitals, however, the final discharging hospital is considered responsible for any readmissions.
 c) From the admission set, cases whose discharge status is "against medical advice." Because in such cases, the planned treatment(s) could not be fulfilled and thus they do not represent a quality-of-care signal.
 d) From the readmission set, those entries that fall within 24 hours of their prior index discharge. This is consistent with the VHA operations policies.
 e) From the readmission set, cases in which any of the CMS planned procedures are conducted if not followed by an acute or a complication-of-care discharge condition category. Examples of such procedures include peripheral vascular bypass, heart valve, kidney transplant, mastectomy, colorectal resection, and maintenance chemotherapy (see [20] for the full list).
 f) From the readmission set, AMI patients hospitalized for a percutaneous coronary intervention (PCI) or coronary artery bypass graft (CABG), except those that are diagnosed for heart failure, AMI, unstable angina, arrhythmia, and cardiac arrest.
 g) From both admission and readmission sets, hospitalizations in long-term care, palliative care, nursing home, aftercare of convalescence, psychiatry, rehabilitation, and hospice wards; or for fitting of prostheses and adjustment devices.
 h) From both admission and readmission sets, stays for special conditions with high mortality risk, for which chances of post-discharge death is much higher than chances of being readmitted. These include, but are not limited to, patients with malignant neoplasm without specification of site; and medical patients with cancers of breast, skin, colon, upper digestive tract, lung, liver, pancreas, head, neck, brain, and fracture of neck of femur (hip). This is consistent with the CMS approach.

i) From both admission and readmission sets, records that are related to major or metastatic malignancies, multiple trauma, burns, neonatal, obstetrical, Human Immunodeficiency Virus (HIV), and eye care. The rationale is that these conditions usually require specialized follow-up cares and are often not avoidable. This is consistent with the 3M approach.

j) From both admission and readmission sets, patients not enrolled in the VA and thus lacking sufficient historical data for the 12 months prior to the index admission. The logic is that the information is required to adjust for the case-mix and comorbidities.

k) From both admission and readmission sets, records with inconsistent and/or error components such as age and gender discrepancies, invalid HCC assignment, discharge date that preceded the admission date, disagreements between the patient's VA status and its service-based attribute values, hospitalizations charged for less than $200 or greater than $4 million, and records with distances longer than 3000 miles.

IV. Calculate *eligible admissions* as all records remaining in the admission set. Note that, situations described in a), b), and c), i.e., "death," "transfer," or "against-medical-advice" may happen to both admission and readmission entries.

**Step 2** (labeling PARs)

V. Mark records from the readmission set that have a clinical relationship with their initial admissions as defined by one of the eight following categories:

a) Readmissions for an ambulatory care-sensitive condition as specified by the Agency for Healthcare Research and Quality (AHRQ) [21].

b) Medical readmissions for repeated happening or extension of the reason for the initial (or a closely-related) condition.

c) Medical readmissions for an acute decompensation of a chronic condition that relates back to the care given in the course or immediately after the initial admission (e.g., a return hospitalization for diabetes by an initially diagnosed AMI patient).

d) Medical readmissions for acute medical complications acquired during or soon after the first admission (e.g., a readmission for addressing a urinary tract infection of a patient originally hospitalized for hernia repair).

e) Readmissions for a mental health or substance abuse condition that follows an admission for a non-mental health or non-substance abuse condition.

f) Readmissions for mental health or substance abuse reason following a hospitalization for a mental health or substance abuse reason.

g) Surgical readmissions to deal with repeated happenings or extensions of the condition causing the initial hospitalization (e.g., a readmission for appendectomy surgery of a patient who was initially admitted for abdominal pain and fever).

h) Surgical readmissions to tackle a medical or surgical complication resulting during the initial admission or in the post-discharge course (e.g., a readmission for treating a post-operative wound resulting from an initial hospitalization for a bowel resection).

**Step 3** (clinical panel review)

VI. All exclusions from step 1 and marked PARs in step 2 are reviewed by three physicians, and final decision about the outcomes was made by a majority of vote scheme.

**Step 4** (calculating PAR rate)

VII. Define a PAR series as a sequence of one or more PARs that are all clinically associated with a similar initial admission. In this way, the succeeding PARs are always assessed for having a clinical relationship in reference to the very first admission (which starts the sequence), not with the intermediate PARs. As a result, the total time interval encompassing a PAR series can be larger than 30 days.

VIII. Update the eligible admission set by reclassifying cases in the readmission set that are NOT found to be PARs (i.e., not having clinical relationship with their prior admissions) and at the same time, do not fall in "death", "transfer", or "against-medical-advice" categories.

IX. Calculate PAR rate as $\frac{\# \text{PAR Series}}{\# \text{Eligible Admissions}}$.

It should be noted that, plus using DCG/HCC system, we utilize other sources of information such as clinical visits between admission and readmission, and communication with the patient, patient's family and primary care physician assigned to judge whether the readmission(s) could have been avoidable.

*Predicting potentially avoidable readmission*

Basically there are two types of prediction models applied in readmission studies. The first group, which we call classification models, focus on readmission occurrence and attempt to estimate it by a learning algorithm trained with inputted patient factors and known class labels. A popular example of this class is logistic regression [11]. The second group, which we name timing based models, concentrates on time to readmission and try to relate some of its probability functions to a given set of covariates in parametric or semiparametric schemes. A well-known example of this category is the Cox proportional hazard model [22]. In this paper, we take a hybrid approach and propose a tree based classification method that can model the effect of partially known information (censored observations) into the risk of readmission. The proposed method is also able to directly incorporate patient's history of readmission and risk factors changes over time.

Consistent with the CMS logic [20, page 14], we observe that time-to-readmission curves for the four conditions follow a similar pattern over time: a quick early increase of rate of readmission, followed by a stable and constant rate thereafter. Thus, it is reasonable to assume that time spent until readmission can be stratified into two groups: one for those who quickly return to hospital possibly due to poor quality of *inpatient* care they receive, and the other for those who slowly readmit because of poor quality of post-discharge and *outpatient* follow-up care. Following this, we develop a conceptual framework for the movements of patients after discharge from hospital (see Figure 1). It is assumed that discharged patients travel between two major states (Short Stay and Long Stay) in their community before being returned to the hospital. In other words, patients begin their post discharge period from the Short Stay (SS) group consisting of $m$ sequential transient phases; then they are either readmitted to the hospital at the rate of $\lambda_{SS}$ or move to the Long Stay (LS) group with rate $\lambda_m$. Patients entering in the LS group remain another $r$ transient phases in the community before going back to the hospital at the rate of $\lambda_{LS}$. Therefore, readmission from the short stay group is a marker of poor quality of inpatient care, whereas those from the long stay group represent deficient quality of outpatient care. Note that the rates are not identical within or between the two groups.

The current framework results in a special case of order $m + r$ Coxian phase-type distribution, which is represented by an absorbing continuous time Markov chain (CTMC) with $m + r$ transient states and one absorbing state (Hospital). See [23] to get an overview of phase-type distribution and its applications on modeling healthcare systems. The dynamics of the underlying finite state stochastic process $\{X(t); t \geq 0\}$ is governed by the transition intensity matrix $\boldsymbol{A} = \{\alpha_{hj}\}; h, j \in E, E = \{1,2,\ldots,m+r\}$ as

$$\alpha_{hj}(t) = \lim_{\Delta t \downarrow 0} \frac{P[X(t+\Delta t) = j | X(t) = h]}{\Delta t},$$

$$\alpha_{hh}(t) = -\sum_{h \neq j} \alpha_{hj}(t). \tag{1}$$

Hence, the random variable time to readmission $T$ is equal to the time spent in the above CTMC until absorption in the Hospital state, which is also known as the sojourn time. The probability density function $f$, the survival function $S$, and the $k$-th (non-central) moment of $T$ are expressed by

$$f(t) = \boldsymbol{\pi} \exp(\boldsymbol{Q}t)(-\boldsymbol{Q}\boldsymbol{1}) \tag{2}$$

$$S(t) = \boldsymbol{\pi} \exp(\boldsymbol{Q}t)\boldsymbol{1} \tag{3}$$

$$m(k) = (-1)^k k! \, \boldsymbol{Q}^{(-k)} \boldsymbol{1}, k = 1,2,\ldots \tag{4}$$

where $\boldsymbol{\pi}$ is a row vector row vector of the initial probabilities over the transient states, $\boldsymbol{Q}$ is a $(m+r) \times (m+r)$ transient partition of the intensity matrix, and $\boldsymbol{1}$ represents an $(m+r) \times 1$ column vector of 1's. Based on the transition flow diagram shown, the Coxian phase-type distribution is represented by $PH(\boldsymbol{\pi}, \boldsymbol{Q})$ where $\boldsymbol{\pi} = (1,0,\ldots,0)$ and $\boldsymbol{Q} = \{q_{hj}\}$ is simplified as

$$q_{h,h+1} = \lambda_h; h = 1,2,\ldots,m+r-1,$$

$$q_{h,h} = -\lambda_h; h = \{1,2,\ldots,m-1, m+1,\ldots,m+r-1\}, \tag{5}$$

$$q_{m,m} = -(\lambda_{SS} + \lambda_m), \quad q_{m+r,m+r} = -\lambda_{LS}.$$

It is worth mentioning that the phases within each major state (short stay or long stay) do not carry any practical interpretations, but time spent in each phase follows an exponential distribution. There are a number of approaches to fit a phase-type distribution to empirical time-to-readmission data $t_i, i \in \{1,2,\ldots,N\}$ [23]. Here, we use expectation-maximization algorithm (EMpht program [24]) to maximize the log-likelihood function

$$L = \sum_i \alpha_i \log(f(t_i)) + (1-\alpha_i) \log(S(t_i)), \tag{6}$$

in which $\alpha_i = 1$ if $t_i$ is a complete time for the $i$-th hospitalization, and zero if $t_i$ is a censored case (i.e., no readmission occurs within 30 days after discharge).

Further, to develop a tree-based classification method, we adopt the basic idea of Breiman's random forest algorithm [25] and utilize the phase-type likelihood function as a splitting criterion instead of the traditional Gini index. The proposed approach can be seen as a special type of random survival forest [26], and thus we name it as phase-type survival forest.

- **Splitting criterion**

We use minimization of the weighted average information criterion (WIC) as the splitting criterion [27]. The WIC is calculated as

$$WIC(d) = -2L + d + \left\{ \frac{d\left(\left((\log(N)-1)\log(N)\right)(N-(d+1))^2 + 2N(N+(d+1))\right)}{\left(2N + \left(\log(N)(N-(d+1))\right)\right)(N-(d+1))} \right\}, \tag{7}$$

where $d = 2(m+r) - 1$, is the number of degrees of freedom for phase-type distribution, and $N$ is the total number of sampled records. In this way, at each node of a tree, if covariate $\ell$ has $G$ values breaking the node into partition set $(\ell_1, \ell_2, \ldots, \ell_G)$, the total WIC for the split can be expressed by the sum of singular WICs of every sub-group as $WIC_{\text{full}}(d_{\text{full}}) = \sum_{g=1}^{G} WIC_{\ell_g}(d_{\ell_g})$. Also, the information gain is defined as the improvement made in the WIC after splitting the node like $IG_\ell = (WIC_R(d_R)) - WIC_{\text{full}}(d_{\text{full}})$, where $R$ stands for the node before partition (i.e., the parent node). Beginning from the root node, at every single node, we apply one covariate at a time and record the gain for partitioning by that covariate. Then, we repeat this with other attributes and select a split that minimizes the WIC the most (or yields the largest gain) to recursively partition into child nodes. Also, if no positive gain can be obtained at a node by any possible split, the node becomes a terminal node.

- **Forest development**

Because we allow multiple records per patient in our data, repeated measures and recurrent readmissions are likely. In this case, the bootstrapped samples are dependent and chances of having correlated observations in the in bag training set are high. Consequently, trees grown may be correlated and overfitting is plausible. To alleviate this problem, we force the forest take a bootstrap sample at the patient level rather than at the replicate level, i.e. doing subject specific bootstrap instead of traditional replicate based bootstrap. This way, when a particular patient is chosen at random, all of its replicates (repeated measures) that had the outcome (recurrent events) or did not have the outcome are attached to it. Consistent with the rule of thumb, subject level bootstrapping performed in the algorithm ensures that about 63% of the subjects (rather than replicates) are elected in-bag. As a result, patients with more repeated measures cannot dominate the learning process. The algorithm, phase-type survival forest, is described in Algorithm 1. Similar to the original Breiman's random forest, the out-of-bag (OOB) data (which includes about one third of all patients) is used to get a running unbiased estimate of the classification error. Likewise, we use the same permutation based measure to get a raw importance score for variable [25].

*Data preprocessing*

Since the data contains missing values, noise (e.g., errors and outliers), and inconsistent records, we perform the following preprocessing tasks:

a) In univariate baseline analysis, missing values are imputed with hot-deck method [28]. In predictive model building, the default Breiman's replacement method is employed [25].

b) Extreme records (outliers) are identified and removed by local outlier factor [29].

c) Error records and incorrect data combinations (such as prisoner-of-war status: YES, veteran status: NO) are fixed manually.

d) Variable 'distance' is discretized into three levels (near: < 25 miles; middle: between 25 and 50 miles; far: > 50 miles) by $k$-means clustering. This is done because distance has a multimodal and highly skewed density function.

Following these steps, the number of records is reduced to 6975 with 2813 distinct patients.

## 3 Analyses

We examine the most frequent diagnoses of 30-day readmissions after hospitalization for heart failure, acute myocardial infarction, pneumonia, and COPD. We compare percentages of readmission calculated by our method with those of the 3M and CMS approaches. We performed a series of analyses to investigate the calibration of the proposed prediction method. To do this, we first create and enter two new covariates into the analysis: (1) 'sequence' that shows how many times a given patient is readmitted, and (2) Charlson comorbidity index with the help of comorbidity software [30]. Then we conduct three sensitivity analyses: (1) sensitivity of error rates to the parameters of the phase-type survival forest (i.e., number of trees to grow at each node, and number of variables to randomly split at each node), (2) sensitivity of error rates to cutoff point of continuous covariates, and (3) sensitivity of error rates to class weights (i.e., readmitted class and not readmitted class). Next, we evaluate observed-to-predicted ratios of our prediction model at different readmission risk deciles with the help of calibration curves. Finally we compare the discrimination power of our prediction model with four classification methods found in the literature. To this end, we use different prediction measures including sensitivity, specificity, positive predictive value, and negative predictive value.

## 4 Results

*Potentially avoidable readmission rates*

Using the potentially avoidable readmission (PAR) algorithm and 30-day timeframe for the four conditions, we begin by classifying all records to admissions and readmissions. After removing instances from the admission and readmission sets that meet one or more exclusion criteria (see section III of the PAR algorithm), we initially identify total of 5,476 eligible admissions and 941 readmissions. Of the 941 readmissions, 155 cases are found not clinically related to their prior admissions (see PAR algorithm, section V), form which 31 cases are fitted in either "death," "transfer," or "against-medical-advice" groups and thus be dropped. The remaining 124 readmissions are then reclassified as eligible admissions, resulting in 5,600 eligible admissions. Hence, we end up having 786 PARs, from which 588 examples belong to a PAR series with only one PAR, and 71 match to a PAR series with two or more PARs. Consequently, the total number of unique PAR series becomes 659, and the PAR rate (see section IX of the PAR algorithm) is found to be 11.77 percent. Following the same approach, rates of PAR for heart failure (HF), acute myocardial infarction (AMI), pneumonia (PN), and COPD are 13.26, 12.47, 11.16, and 11.18 percent. The facility adjusted PAR rates vary from 12.37% to 13.69% for HF; 11.83%–13.16% for AMI; 10.74%–11.93% for PN, and 10.47%–12.05% for COPD. From all HF avoidable readmissions, 86.3% are readmitted once, 11.4% are readmitted twice and 2.3% are readmitted three or more times. These rates are (81.7%; 14.6%; 3.7%), (88.4%; 10.9%; 0.7%), and (85.1%; 13.5%; 1.4%) for AMI, PN, and COPD respectively.

The most common diagnoses of 30 day readmission are outlined in Table 1. It appears that after admission for HF and AMI, readmissions happen mostly for heart failure (39.6% and 28.3% of readmissions, respectively), but following hospitalizations for PN and COPD, patients get readmitted because of COPD (21.4% and 31.6%, in turn). Also, the top five readmission diagnoses contribute to 63.2% of all readmissions after HF, 59.4% of all readmissions after AMI, 55.6% of all readmissions after PN, and 62.4% of all readmissions after COPD. Also we observe that the most frequent reasons for avoidable readmissions in all conditions are related to "recurrence or extension of the reason (Section V, part b)" and "medical complications (Section V, part d)", with an average of 54.7% and 23.2% through all the hospitals. As expected, in none of the acute and chronic conditions is the proportion of non-clinically related readmissions over 15.4 percent.

We compare percentages of readmissions calculated by our method (PAR) to those of the 3M method for the three acute conditions in the four hospitals (Figure 2). It is noticed that with our approach (PAR), a greater proportion of all readmissions can be avoided in the first two weeks after discharge, but the contribution declines as time passes. Compared to the 3M approach, our method considers (slightly) fewer rehospitalizations as being avoidable and produces lower rates of readmission throughout all periods after discharge. A probable reason for this may be related to the CMS and VHA specific exclusions of our method, which is not found in the 3M approach.

*Descriptive Analytics*

Turning to the description of the underlying population, we observe that the mean (standard deviation) patient age of the readmitted cohort is 78.6 years (3.5 years) for HF, 80.3 years (4.1 years) for AMI, 79.3 years (2.9 years) for PN, and 77.1 years (2.9 years) for COPD. Frequent comorbid conditions among readmissions are coronary artery disease (CAD), atrial fibrillation, and diabetes for the HF cohort; anemia, congestive heart failure, and vascular disease with complications for the AMI cohort; chronic obstructive pulmonary disease, congestive heart failure, and cardiorespiratory failure and shock for the PN cohort; and chronic bronchitis, pneumonia, and diabetes mellitus for the COPD cohort.

Baseline patient characteristics in cohorts with potentially avoidable readmission (PAR) and without any kind of readmission (No readmission) are displayed in Table 2 (for heart failure and acute myocardial infarction) and Table 3 (for pneumonia and COPD). The presence of any significant difference between the cohorts is also tested using univariate logistic regression and the results are shown in terms of $P$ values (with missing values imputed by the hot-deck method). Since the same patient could have several avoidable readmissions during the study period, we used generalized estimation equation to adjust for serial correlations among readmissions of the same patient.

During the study, a total of 5,600 eligible admissions were made in the four VA hospitals, out of which about 13.09% were followed by an unnecessary rehospitalization. Note that this rate is different from what is reported before (which is 11.77%) because here we count each readmission separately rather than as members of a PAR series. . In all conditions, the populations are generally male (>86%), married (>51%), older (>67 years), and live within 25 miles of a VA facility (>60%). More than 21% in all conditions do not have private insurance or insurance through Medicare or Medicaid programs. More than half of patients in all conditions are admitted directly from their home and more than 50% have one to four past year hospitalizations. On average, the care assessment score is higher in respiratory diseases (near 69) compared to circulatory conditions (about 66). Almost 18% of the patients are also diagnosed with more than ten HCCs (not shown in the tables). Note that in the attribute "source of admission," class 'transfer' is related to those patients who are transferred only among the four VA hospitals, and 'Other' is related to some other admission sources such as observation/examination, non-VA hospitals not under VA auspices, community nursing homes under (or not under) VA auspices, non-veteran hospitals, etc. Priority groups 1, 2, and 3 are generally assigned to veterans with service connected disabilities of $> 50\%$, $[30\%, 50\%)$, and $[20\%, 30\%)$, respectively. Other groups are as follows: 4, catastrophically disabled veterans; 5, low income or Medicaid; 6, Agent Orange or Gulf War veterans; 7, non-service connected with income being below HUD; and 8, non-service connected with income being above HUD. For each condition, patient comorbidities are identified with the help of comorbidity software, using ICD-9-CM and DRG codes from the index hospitalization and any admission in the 12 months prior.

It is observed that patients who are subsequently readmitted are elderly and usually have a greater number of comorbidities. Male patients have on average a greater chance to be readmitted in HF and COPD cohorts rather than females, but this cannot be generalized since the VA sample here contains only about 8% female patients. The analysis shows that length of stay is not generally associated with odds of avoidable readmission, when patient and facility characteristics are not controlled for. However, after adjusting for the case mix and service mix (not shown here), the relation tends to be negative (about 7.3% increase for each in hospital day lower than expected), which implies that shorter individual length of stay is generally connected with higher risk of readmission. Therefore, consistent with [31], we observe that significant reduction in LOS, without simultaneously improving inpatient care, is more likely to result in premature discharge and rehospitalization. Enrollment priority turns out to be highly linked with odds of readmission in all conditions, especially when it comes to catastrophically disabled veterans (increases of .2% in AMI to 10.9% in HF). Also the odds of avoidable readmissions are significantly higher in patients exposed to ionizing radiation and Agent Orange in all conditions. Among the comorbid conditions, having diabetes and cancer increases the chance of readmission, as does having mental disorders and substance abuse (with harsher effect in circulatory conditions).

*Predictive Analytics*

Following Algorithm 1, we used the entire set of patient risk factors to develop a readmission prediction model. For non-categorical variables in the candidate set (i.e., age, length of stay, CAN score, sequence, and Charlson index), we evaluated different cut off points to split the dataset into binary partitions and explore the optimal cut-point that most discriminates high vs. low risk using operating characteristic curves (ROC) .We then used these cut-points for further analyses. Also for categorical features with more than two classes (like race), following the literature, we optimally select a series of binary splits (instead of multiway splits) that produce the best discrimination results.

We first begin with the baseline model that uses all sampled data points and we let the forest internally perform cross validation using out-of-bag (OOB) samples during each run. The number of trees and the number of variables to try at each split are set to 6,000 and 5, respectively. Also we set the cut-points with respect to minimizing the WIC criterion as follows: age, 68 (years); length of stay, 5 (days); CAN score, 66; sequence, 3; and Charlson index, 4.5. Results of variable importance are summarized in Table 4 (Sig. stands for significance level). As illustrated, almost all statistically significant variables (Sig. <.05) refer to overall health and agedness factors, which may reflect a generalized vulnerability to disease among recently discharged patients—inpatients regularly lose their strength and develop new difficulties in doing their day to day activities. Interestingly, 'sequence' turns out to be (positively) related to readmission risk, which highlights the fact that the chance of unnecessary returns to hospital is greater in patients with prior history of readmission. In the baseline model, the $c$-statistics is .793; sample-level OOB error rates are 3.16%, 2.35%, and 8.05% for overall, No readmission class, and PAR class, respectively; and there are large interactions between Agent Orange and Radiation, between Priority and Length of stay, and between Priority and Insurance, to name a few.

- **Model calibration**

We then calibrated the baseline model as follows: (1) we focused only on the 16 most important variables found in the baseline model; (2) we imputed missing values based on Breiman's replacement method [25]; (3) we modified the optimal cut off points with regards to maximizing the $c$-statistics (the new cut-points are 69 years for age, 70 for CAN score, and 4.7 for Charlson index, while others remain unchanged); and 4) we altered the class weights to 1 on class 'No readmission' and 8 on class 'PAR', to adjust for the imbalanced prediction errors in the classes. Then we rerun the model with 10,000 trees and 4 variables to try at each split. Depiction of variable importance for the calibrated model is shown in Table 5. Expectedly, the ranking of variables does not change but we achieved better results in terms of scores and significance levels. It is noticed that, though Mental disorder and Malignant neoplasm are only marginally significant, we decide to keep them in the final model since 1) they are both medically significant in contribution to the risk of readmission, and 2) they together contribute largely to the model discrimination ability.

In the calibrated model, the $c$-statistics jumps to .836; no serious interactions remain among variables; and the overall, No readmission, and PAR error rates become 3.67%, 2.51%, and 2.64%, respectively. It is remarkable that the calibrated model considerably decreases PAR misclassification rate, but at the expense of increasing the overall error rate a little bit. We perceive that this tuning in class weights is really appealing for our situation because in readmission prediction models, the cost of false negatives (which correspond to readmitted patients incorrectly predicted as No readmission) is usually much higher than the cost of false positives (which correspond to non-readmitted patients incorrectly predicted as PAR cases).

We further check the calibration by evaluating predicted and actual PAR rates at different risk deciles. These results appear in Table 6 and Figure 3. It is noted that both on average and over the whole range of predictions, the predicted probability of readmission matches up well with the actual probabilities. Average predicted readmission (not shown here) also monotonically increases with growing risk, ranging from 8.79% in the lowest decile to 43.75% in the highest, a range of 34.96% in total. For the 12% of readmissions that happens between deciles four and five, the proposed model under predicts by roughly 8.5%. It also over predicts by about 4%–14% for the small number of readmissions (21%) which occur in deciles 6–10.

- **Model validation**

We used the calibrated model and studied its internal validity (also called reproducibility), based on the same population underlying the sample. To this end, since the proposed method does perform bootstrapping internally, we slightly modified the split sample technique for our purposes: we randomly partitioned the sample into 50% training and 50% testing sets and redid this 7 times. For each partition we ran the proposed algorithm and obtained the $c$-statistics. The average $c$-statistics for the seven runs of training sets reached .839 and for the test sets, it was .821. Hence, there exists an "optimism" of .018 in the mean area under the ROCs for the training and testing splits, and as a result, the internally validated (or optimism corrected) $c$-statistics is estimated as .818.

To provide more robust evidence of validity, we further conducted external (in fact: spatial) validation (also called generalizability) with a new sample of 478 patients admitted (with primary diagnosis of heart failure, acute myocardial infarction, pneumonia, and COPD) in the months of August and September 2012. It is noted that we included the same patient factors studied in the new sample. The $c$-statistics in the external sample decreased to .809 (a decrease of .027) which is slightly more than results from internal validation (a decrease of .018). However, both internal and external validations confirm the superiority of our proposal over the current approaches in terms of discrimination power and stability. Nonetheless, we obtain greater c statistics (at least .813) when the proposed method is applied separately on each condition. It should also be remarked that with the current sample data, the CMS endorsed model can only produce a c statistics of about .63.

*Comparisons with other approaches*

We evaluate our proposed method (PHSF: phase-type random forest) with logistic regression, Breiman's Radom Forest, Support Vector Machine (SVM), and neural network in terms of different predictive measures such as sensitivity, specificity, positive predictive value (PPV), negative predictive value (NPV), F-score (which can be interpreted as a harmonic mean of sensitivity and PPV), Matthews correlation coefficient (MCC), mean square error (MSE), and area under ROC curve (AUROC) [32]. The models are built and compared with the R version 3.0.2 [33] using packages `randomForest` [34], `e1071` [35], `glm2` [36], and also MATLAB neural network toolbox [37]. It is worth mentioning that we used different kernels such as polynomial and radial basis function for the SVM method. For the neural network approach, we tested for two and three layers with different numbers of sigmoid hidden neurons and linear output neurons. For the pure random forest method, we did the same calibration as with the proposed method, and for the logistic regression, we used generalized estimation equation to account for clustering at the patient level. The comparison results are summarized in Table 7 and Figure 4. As shown, the proposal works better than other alternatives in all predictive criteria. The Breiman's random forest approach and SVM produce very close results in this sample but the neural network approach seems unable to compete with other models having a modest discrimination of about 0.7. Not surprisingly, all models predict 'No readmission' cases better than the PAR cases. It is of interest that SVM slightly outperforms the random forest in terms of PPV (.5% higher) and true negative rate (.17% higher). Furthermore, in the overall spectrum of false positive rates, the proposal assigns a higher probability of readmission for a patient with PAR compared to a 'No-readmission' patient, about 83.6% of the times. Looking at different ROC stairs graphs, we can infer that, with a false positive rate varying between .09 to .25, our PHSF approach is placed higher than the others, but it falls behind the SVM and neural network in case of very small rates of false positive. In higher false positive rates, we observe that random forest and SVM are very similar in discrimination ability and they work as well as our proposal beyond .7 false positive rate. However, logistic regression turns out to fall short at a false positive rate of .8 to .9.

# 5 Conclusion

Concentration on reducing unnecessary readmission has never been higher, especially with the CMS augmenting the rates of penalties and introducing new waves of diseases that will be under scrutiny during next years. In response to this policy shift, hospitals and clinicians are become more interested in analytics ways to identify patients at elevated risk of avoidable readmission, since such tools can ultimately be used to guide more appropriate discharge planning and efficient resource utilization. Although a variety of approaches have been proposed to identify patients with higher risk, their potentials have been limited mainly because they do not incorporate timing of readmission in their prediction and/or they are not accurate enough [7].

In this study, we make several contributions to readmission reduction studies. First, we address the problem of characterizing avoidable (or unnecessary) readmissions from all other types of outcomes. Our algorithm (PAR) is based on administrative data and takes a more accurate look at preventability components of rehospitalization compared to existing methods. We also suggest using a more comprehensive risk adjustment tool (DCG/HCC) in counting avoidable readmissions, as well as getting help from other sources of information, like clinic visits between index admission and readmission, in assessing the avoidability of readmissions. Second, we propose a hybrid prediction model that exploits good aspects of classification and timing based analytics models. We then demonstrate the superiority of our model over current solutions with respect to various accuracy criteria. Further, to confirm that the high discrimination ability of our proposal is irrespective to overfitting, we perform internal and external validation practices. Also, unlike some studies in the literature, we do not limit our work to a specific disease or within a specific hospital, but instead we aggregate data from four different VA facilities containing inpatients diagnosed with four different conditions.

Even though our results introduce new aspects of readmission studies, one should pay attention to some limitations in interpreting and generalizing them. First, the data used in the study is from one region (Veteran Integrated Service Network 11, Veterans In Partnership) in the State of Michigan, with a veteran population that is mostly male and veteran, and a government funded care delivery system; hence the results may not be identical in other health care systems. Second, the study is limited to administrative data (that are regularly available to all health plans) and it does not have laboratory test results and vital signs such as hemoglobin or serum level at discharge, which may affect the risk of unnecessary readmission.

In future work, we plan to use our proposal to compare and profile the hospitals on their readmission rates using proper risk adjustment for case mix and service mix. The approach currently employed by the CMS (and the VHA) is to calculate a ratio of observed to expected outcomes for a given hospital, and evaluate it across the normal range of all other hospitals given the same mix. Methods in this context are primarily based on models in which the hospital effects on outcome are taken as random. Nonetheless, they have been recently argued because 1) they often produce biased estimates of outcomes at the provider level; and 2) they cannot prevent confounding issues when the patient characteristics are correlated with facility effects [38].

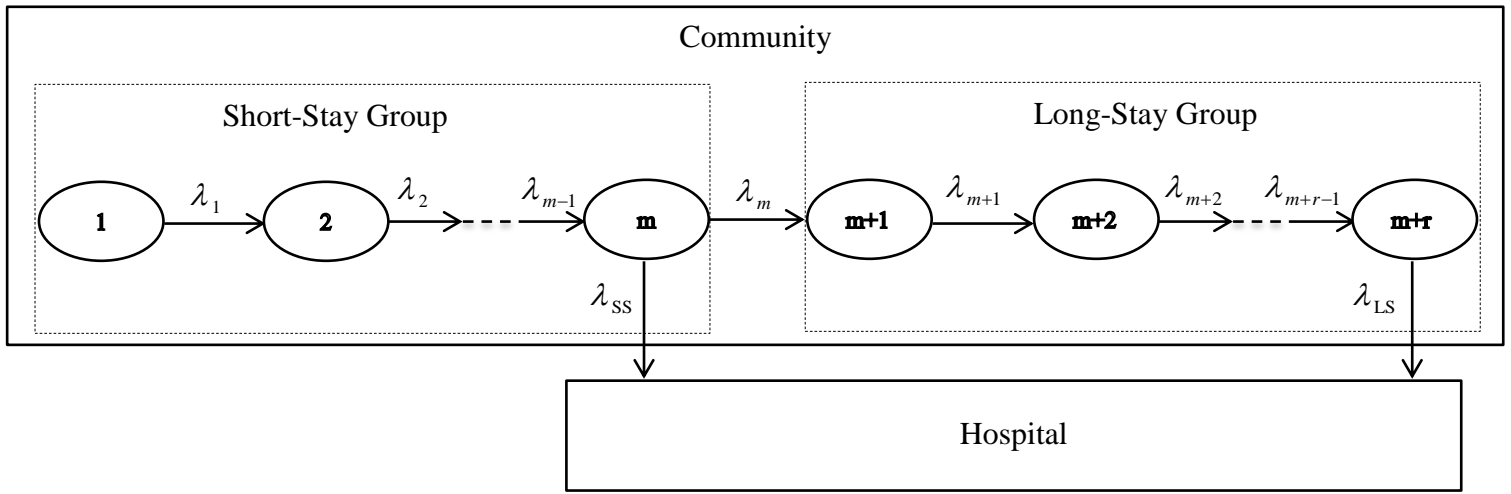

**Fig.1** Markov model for movements of patients after discharge

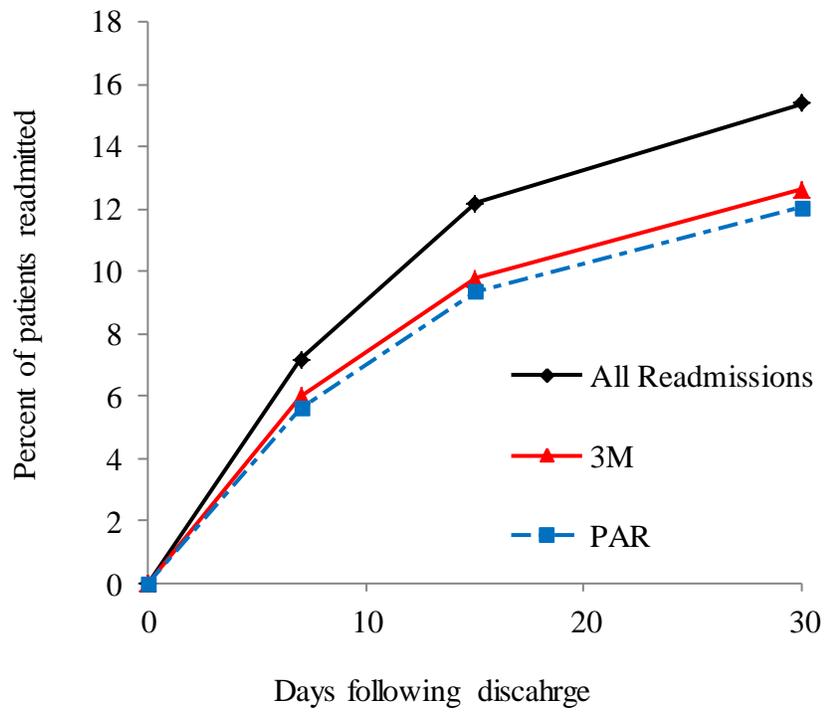

**Fig. 2** Percent of readmission over time

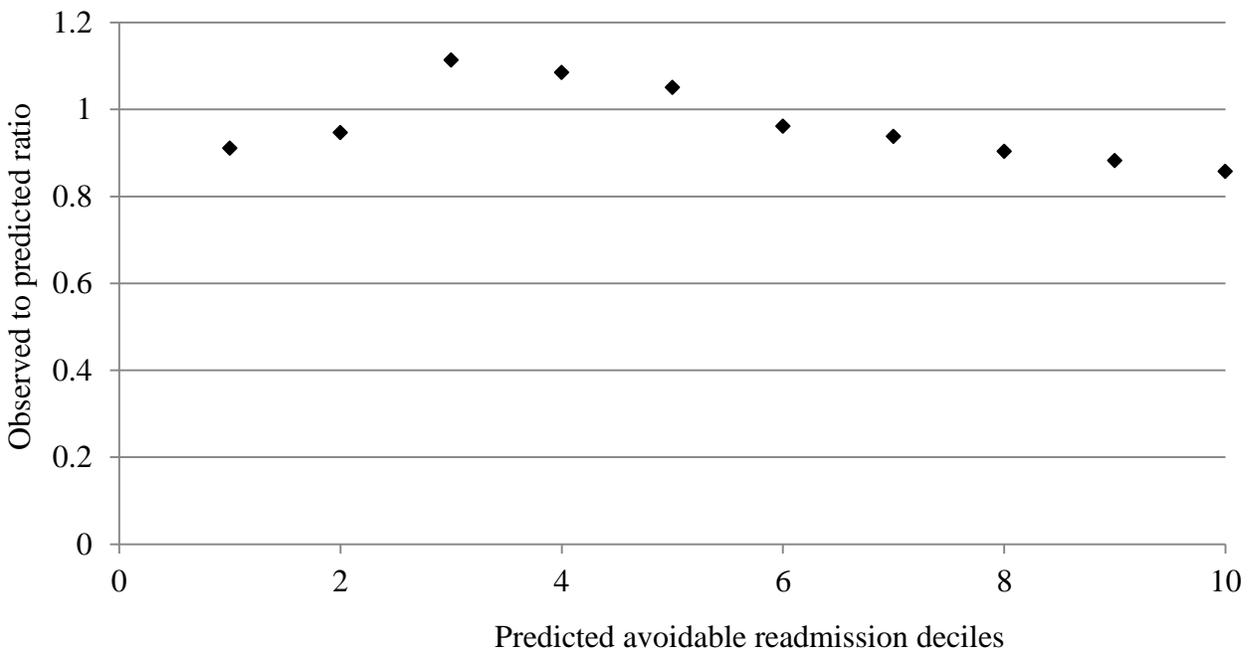

**Fig. 3** Calibration curve for the proposed model

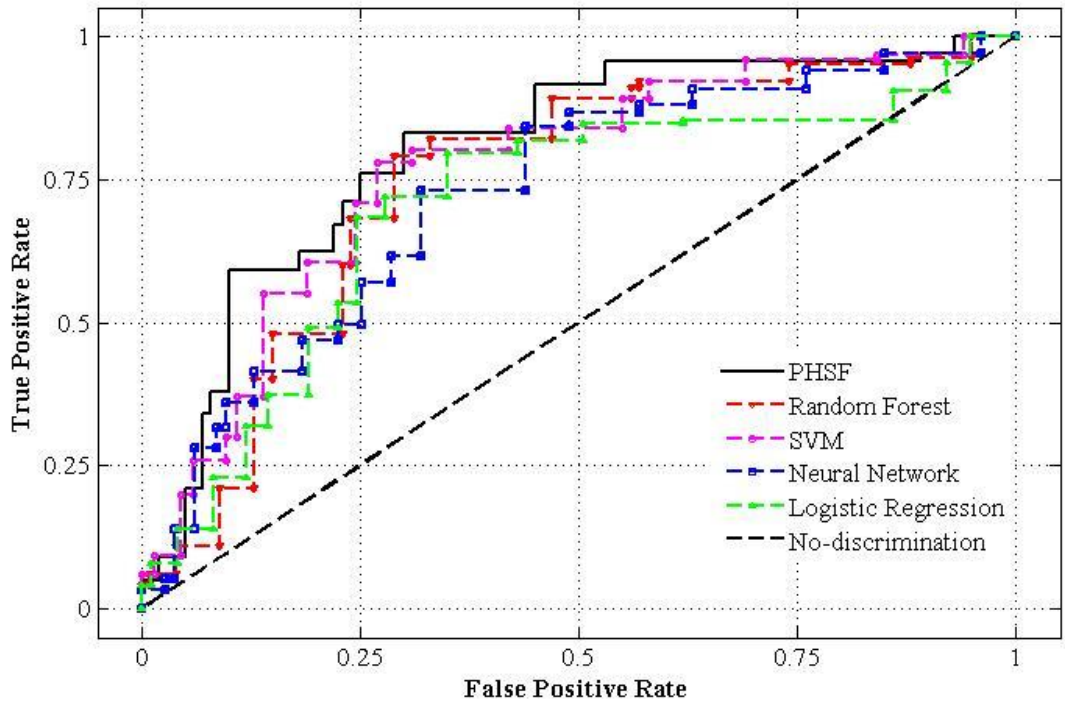

**Fig. 4** ROC curves for different predictive models

**Table 1** Top readmission diagnoses for patients hospitalized after heart failure, acute myocardial infarction, pneumonia, and COPD

| Rank | HF cohort | | AMI cohort | | PN cohort | | COPD cohort | |
|---|---|---|---|---|---|---|---|---|
| | Diagnosis | Percent of PAR | Diagnosis | Percent of PAR | Diagnosis | Percent of PAR | Diagnosis | Percent of PAR |
| 1 | Heart failure | 39.6% | Heart failure | 28.3% | COPD | 21.4% | COPD | 31.6% |
| 2 | Renal failure | 9.3% | Coronary artery disease | 13.7% | Pneumonia | 15.3% | Bronchitis | 15.8% |
| 3 | Arrhythmias | 6.7% | Pneumonia | 8.6% | Heart failure | 10.6% | Cardio-Respiratory Failure and Shock | 7.2% |
| 4 | Cardio-respiratory failure and shock | 4.1% | Septicemia/ Shock | 5.5% | Cardio-respiratory failure and shock | 4.4% | Pneumonia | 4.6% |
| 5 | Pneumonia | 3.5% | Renal failure | 3.3% | Renal failure | 3.9% | Hypertension | 3.2% |

**Table 2** Baseline characteristics (mean (SD) for continuous variables; n(%) for categorical variables) and univariate analyses at the time of discharge

| Characteristic | Heart Failure (n=1674) | | | Acute Myocardial Infarction (n=1417) | | |
|---|---|---|---|---|---|---|
| | No Readmission (n=1447) | PAR (n=227) | P-Value | No Readmission (n=1211) | PAR (n=206) | P-Value |
| Age (years) | 68.6 (5.2) | 71.3 (3.2) | <.01 | 69.3 (5.6) | 73.3 (3.7) | <.01 |
| Sex, Male | 1406 (97.2) | 215 (96.9) | .04 | 1097 (90.6) | 192 (93.2) | .07 |
| Race | | | | | | |
|   Black | 986 (68.1) | 193 (85.0) | | 769 (63.5) | 169 (82.0) | |
|   White | 432 (29.8) | 29 (12.8) | <.01 | 405 (29.8) | 29 (14.1) | <.01 |
|   Other | 29 (2.1) | 5 (2.2) | | 37 (3.1) | 8 (3.9) | |
| Marital status | | | | | | |
|   Current spouse | 839 (57.9) | 137 (58.3) | | 631 (52.1) | 112 (54.4) | |
|   Never married | 307 (21.2) | 52 (21.4) | .35 | 320 (26.4) | 58 (26.7) | .42 |
|   Previously married | 301 (20.9) | 38 (20.3) | | 260 (21.5) | 36 (18.9) | |
| Primary insurance | | | | | | |
|   Medicare | 732 (50.6) | 126 (55.5) | | 624 (51.5) | 97 (47.1) | |
|   Medicaid | 249 (17.2) | 27 (11.9) | .03 | 226 (18.7) | 32 (15.5) | .07 |
|   Private | 107 (7.4) | 25 (11.0) | | 103 (8.5) | 28 (13.6) | |
|   Not insured | 359 (24.8) | 49 (21.6) | | 258 (21.3) | 49 (23.8) | |
| Length of stay (days) | 5.2 (6.1) | 6.2 (4.4) | .07 | 5.8 (5.8) | 5.1 (6.8) | .11 |
| Source of admission | | | | | | |
|   Direct from home | 797 (55.1) | 129 (56.8) | | 623 (51.4) | 107 (51.9) | |
|   Outpatient clinic | 392 (27.1) | 63 (27.8) | | 392 (32.4) | 67 (32.5) | |
|   Transfer | 17 (1.2) | 3 (1.3) | .31 | 23 (1.9) | 4 (1.9) | .26 |
|   VA NHCU | 62 (4.3) | 12 (5.3) | | 62 (5.1) | 10 (4.9) | |
|   VA Domiciliary | 13 (0.9) | 4 (1.8) | | 13 (1.1) | 5 (2.4) | |
|   Other | 166 (11.5) | 16 (7.0) | | 98 (8.1) | 13 (6.3) | |
| Enrollment priority | | | | | | |
|   1 | 126 (8.7) | 17 (7.5) | | 104 (8.6) | 19 (9.2) | |
|   2 | 167 (11.5) | 9 (4.0) | | 136 (11.2) | 13 (6.3) | |
|   3 | 293 (20.2) | 38 (16.7) | | 239 (19.7) | 41 (19.9) | |
|   4 | 173 (12.0) | 52 (22.9) | | 133 (11.0) | 23 (11.2) | |
|   5 | 316 (21.8) | 66 (29.1) | <.001 | 331 (27.3) | 56 (27.2) | <.001 |
|   6 | 115 (7.9) | 15 (6.6) | | 172 (14.2) | 12 (5.8) | |
|   7 | 103 (7.1) | 19 (8.4) | | 26 (2.1) | 15 (7.3) | |
|   8 | 154 (10.6) | 11 (4.8) | | 70 (5.8) | 27 (13.1) | |
| Distance to hospital | | | | | | |
|   Near (<25m) | 856 (59.2) | 155 (68.3) | | 781 (64.5) | 151 (73.3) | |
|   Middle ([25, 50]m) | 549 (37.9) | 69 (30.4) | .02 | 406 (33.5) | 53 (25.7) | .03 |
|   Far (>50m) | 42 (2.9) | 3 (1.3) | | 24 (2.0) | 2 (1.0) | |
| Prisoner of war, Yes | 17 (1.2) | 8 (3.5) | <.01 | 11 (0.9) | 6 (2.8) | .01 |

**Table 2—Continued** Baseline characteristics (mean (SD) for continuous variables; n(%) for categorical variables) and univariate analyses at the time of discharge

| Characteristic | Heart Failure (n=1674) | | | Acute Myocardial Infarction (n=1417) | | |
|---|---|---|---|---|---|---|
| | No Readmission (n=1447) | PAR (n=227) | P-Value | No Readmission (n=1211) | PAR (n=206) | P-Value |
| Radiation, Yes | 11 (0.8) | 5 (2.2) | .03 | 9 (0.7) | 6 (2.9) | .02 |
| Agent Orange, Yes | 63 (4.4) | 16 (7.0) | .02 | 42 (3.5) | 13 (6.3) | .03 |
| CAN score | 67.4 (4.1) | 71.7 (2.9) | <.01 | 64.5 (4.6) | 68.6 (3.7) | .02 |
| No. of past year hospitalization | | | | | | |
|   0 | 663 (45.8) | 71 (31.3) ⎤ | | 503 (41.5) | 52 (25.2) ⎤ | |
|   1-4 | 713 (49.3) | 122 (53.7) ⎥ | <.001 | 616 (50.9) | 124 (60.2) ⎥ | <.001 |
|   >4 | 71 (4.9) | 34 (15.0) ⎦ | | 92 (7.6) | 30 (14.6) ⎦ | |
| Comorbidity | | | | | | |
|   CAD | 486 (33.6) | 94 (41.4) | .04 | 81 (6.7) | 16 (7.8) | .53 |
|   Heart failure | — | — | — | 346 (28.6) | 73 (35.4) | .04 |
|   Vascular disease w/c | 202 (14.0) | 45 (19.8) | .02 | 306 (25.3) | 67 (32.5) | .02 |
|   Cardiorespiratory | 153 (10.6) | 37 (16.3) | .01 | 134 (11.1) | 14 (6.8) | .06 |
|   Pneumonia | 97 (6.7) | 19 (8.4) | .32 | 51 (4.2) | 15 (7.3) | .05 |
|   Atrial fibrillation | 403 (27.9) | 77 (33.9) | .05 | 291 (24.0) | 62 (30.1) | .04 |
|   Anemia | 225 (15.5) | 47 (20.7) | .05 | 378 (31.2) | 81 (39.3) | .03 |
|   Diabetes | 351 (24.3) | 71 (31.3) | .02 | 159 (13.1) | 37 (18.0) | .05 |
|   COPD | 242 (16.7) | 49 (21.6) | .05 | 63 (5.2) | 17 (8.3) | .07 |
|   Chronic bronchitis | 83 (5.7) | 12 (5.3) | .66 | 17 (1.4) | 6 (2.9) | .14 |
|   Malignant neoplasm | 71 (4.9) | 19 (8.4) | .03 | 25 (2.1) | 12 (5.8) | <.01 |
|   Mental disorder | 160 (11.1) | 37 (16.3) | .01 | 102 (8.4) | 31 (10.7) | <.01 |
|   Substance abuse | 118 (8.2) | 31 (13.7) | <.01 | 112 (9.2) | 33 (16.0) | <.01 |

**Table 3** Baseline characteristics (mean (SD) for continuous variables; n(%) for categorical variables) and univariate analyses at the time of discharge

| Characteristic | Pneumonia (n=1306) | | | COPD (n=1203) | | |
|---|---|---|---|---|---|---|
| | No Readmission (n=1117) | PAR (n=189) | P-Value | No Readmission (n=1039) | PAR (n=164) | P-Value |
| Age (years) | 67.7 (4.9) | 68.3 (2.8) | <.01 | 64.2 (4.4) | 65.1 (2.7) | <.01 |
| Sex, Male | 1035 (92.7) | 182 (96.3) | .07 | 977 (94.0) | 160 (97.6) | .04 |
| Race | | | | | | |
|   Black | 731 (65.4) | 153 (81.0) | <.01 | 601 (57.8) | 121 (73.8) | <.01 |
|   White | 335 (30.0) | 25 (13.2) | | 398 (38.3) | 38 (23.2) | |
|   Other | 51 (4.6) | 11 (5.8) | | 40 (3.9) | 5 (3.0) | |
| Marital status | | | | | | |
|   Current spouse | 571 (51.1) | 106 (56.1) | .27 | 582 (56.0) | 99 (60.4) | .08 |
|   Never married | 244 (21.8) | 32 (16.9) | | 208 (20.0) | 38 (23.1) | |
|   Previously married | 302 (27.1) | 51 (27.0) | | 249 (24.0) | 27 (16.5) | |
| Primary insurance | | | | | | |
|   Medicare | 602 (53.9) | 91 (48.1) | .06 | 541 (52.1) | 98 (59.8) | .03 |
|   Medicaid | 185 (16.6) | 24 (12.7) | | 159 (15.3) | 15 (9.1) | |
|   Private | 89 (8.0) | 26 (13.8) | | 95 (9.1) | 8 (4.8) | |
|   Not insured | 241 (21.6) | 48 (25.4) | | 244 (23.5) | 43 (26.3) | |
| Length of stay (days) | 4.9 (5.4) | 5.7 (4.2) | .03 | 3.8 (5.0) | 4.2 (3.2) | .08 |
| Source of admission | | | | | | |
|   Direct from home | 651 (58.3) | 114 (60.3) | .09 | 581 (55.9) | 98 (59.7) | .39 |
|   Outpatient clinic | 225 (20.1) | 40 (21.2) | | 328 (31.6) | 47 (28.7) | |
|   Transfer | 21 (1.9) | 5 (2.6) | | 33 (3.2) | 3 (1.8) | |
|   VA NHCU | 59 (5.3) | 14 (7.4) | | 62 (6.0) | 9 (5.6) | |
|   VA Domiciliary | 16 (1.4) | 5 (2.6) | | 17 (1.6) | 1 (0.6) | |
|   Other | 145 (13.0) | 11 (5.8) | | 18 (1.7) | 6 (3.6) | |
| Enrollment priority | | | | | | |
|   1 | 74 (6.6) | 22 (11.6) | <.001 | 119 (11.4) | 25 (15.3) | .01 |
|   2 | 141 (12.6) | 17 (9.0) | | 51 (4.9) | 14 (8.6) | |
|   3 | 219 (19.6) | 35 (18.5) | | 182 (17.5) | 21 (12.8) | |
|   4 | 115 (10.3) | 29 (15.3) | | 206 (19.8) | 38 (23.2) | |
|   5 | 341 (30.5) | 36 (19.0) | | 348 (33.5) | 50 (30.5) | |
|   6 | 172 (15.4) | 8 (4.2) | | 22 (2.2) | 5 (3.0) | |
|   7 | 37 (3.3) | 14 (7.4) | | 27 (2.6) | 6 (3.6) | |
|   8 | 18 (1.6) | 28 (7.4) | | 84 (8.1) | 5 (3.0) | |
| Distance to hospital | | | | | | |
|   Near | 692 (62.0) | 127 (67.2) | .01 | 721 (69.4) | 125 (76.2) | <.01 |
|   Middle | 421 (37.7) | 59 (31.2) | | 312 (30.0) | 35 (21.3) | |
|   Far | 4 (0.4) | 3 (1.6) | | 6 (0.6) | 4 (2.5) | |
| Prisoner of war, Yes | 23 (2.1) | 11 (5.8) | <.01 | 27 (2.6) | 14 (8.5) | <.01 |

**Table 3—Continued** Baseline characteristics (mean (SD) for continuous variables; n(%) for categorical variables) and univariate analyses at the time of discharge

| Characteristic | Pneumonia (n=1306) | | | COPD (n=1203) | | |
|---|---|---|---|---|---|---|
| | No Readmission (n=1117) | PAR (n=189) | *P*-Value | No Readmission (n=1039) | PAR (n=164) | *P*-Value |
| Radiation, Yes | 10 (0.9) | 8 (4.2) | <.01 | 13 (1.2) | 9 (5.5) | <.01 |
| Agent Orange, Yes | 39 (3.5) | 15 (7.9) | <.001 | 64 (6.2) | 21 (12.8) | <.001 |
| CAN score | 68.3 (4.6) | 69.1 (2.8) | <.01 | 71.1 (3.6) | 73.2 (2.6) | <.01 |
| No. of past year hospitalization | | | | | | |
|   0 | 485 (43.4) | 56 (29.6) | | 537 (51.7) | 27 (16.5) | |
|   1–4 | 593 (53.1) | 114 (60.3) | <.01 | 449 (43.2) | 114 (69.5) | <.001 |
|   >4 | 39 (3.5) | 19 (10.1) | | 53 (5.1) | 23 (14.0) | |
| Comorbidity | | | | | | |
|   CAD | 216 (19.3) | 31 (16.4) | .3 | 141 (13.6) | 23 (14.0) | .69 |
|   Heart failure | 335 (27.7) | 71 (34.5) | .03 | 131 (12.6) | 15 (9.1) | .46 |
|   Vascular disease w/c | 181 (16.2) | 35 (18.5) | .4 | 91 (8.7) | 11 (6.7) | .19 |
|   Cardiorespiratory | 273 (24.4) | 58 (30.7) | .05 | 106 (10.2) | 10 (6.1) | .10 |
|   Pneumonia | — | — | — | 366 (35.2) | 66 (40.2) | .09 |
|   Atrial fibrillation | 66 (5.7) | 14 (7.4) | .3 | 39 (3.7) | 6 (3.6) | .31 |
|   Anemia | 33 (3.0) | 10 (5.3) | .09 | 19 (1.8) | 4 (2.4) | .11 |
|   Diabetes | 132 (11.8) | 35 (18.5) | .01 | 291 (28.0) | 54 (32.9) | .05 |
|   COPD | 339 (30.3) | 69 (36.5) | .04 | — | — | — |
|   Chronic bronchitis | 72 (6.4) | 9 (4.8) | .4 | 409 (39.4) | 81 (49.4) | <.01 |
|   Malignant neoplasm | 31 (3.1) | 10 (5.3) | .06 | 158 (15.2) | 42 (25.6) | <.001 |
|   Mental disorder | 106 (9.5) | 27 (14.3) | .03 | 233 (22.4) | 47 (28.6) | .04 |
|   Substance abuse | 138 (12.4) | 33 (17.5) | .04 | 272 (26.2) | 55 (33.5) | .03 |

**Table 4** Variable importance for the baseline model

| Attribute | Raw score | Z-score | Sig. |
|---|---|---|---|
| Care Assessment Need (CAN) score | 4.87 | 2.372 | .009 |
| Age | 4.53 | 2.296 | .011 |
| Charlson Comorbidity Index | 4.17 | 2.010 | .022 |
| No. of past-year hospitalization | 4.09 | 1.816 | .035 |
| Sequence | 3.85 | 1.738 | .041 |
| Length of stay | 3.79 | 1.658 | .049 |
| Coronary artery disease | 3.36 | 1.390 | .082 |
| Vascular disease w/c | 3.41 | 1.381 | .084 |
| Admission source | 3.21 | 1.303 | .096 |
| Atrial fibrillation | 3.28 | 1.255 | .105 |
| Priority | 2.88 | 1.068 | .143 |
| Agent Orange | 2.52 | .961 | .168 |
| Pneumonia | 2.75 | .930 | .176 |
| Sex | 2.19 | .869 | .194 |
| Mental disorder | 2.66 | .815 | .207 |
| Malignant neoplasm | 2.53 | .762 | .223 |
| Race | 1.55 | .653 | .257 |
| Radiation | 1.43 | .564 | .286 |
| Cardiorespiratory disease | 1.71 | .550 | .291 |
| Insurance | 1.21 | .483 | .314 |
| Heart failure | 1.17 | .466 | .321 |
| Diabetes | 1.64 | .454 | .325 |
| Prisoner of war | .88 | .330 | .371 |
| COPD | 1.42 | .323 | .373 |
| Marital status | .80 | .283 | .389 |
| All others | .63 | .197 | .422 |

**Table 5** Variable importance for the calibrated model

| Attribute | Raw score | Z-score | Sig. |
|---|---|---|---|
| Care Assessment Need (CAN) score | 7.88 | 3.582 | <.0001 |
| Age | 7.32 | 2.874 | .002 |
| Charlson Comorbidity Index | 7.06 | 2.398 | .008 |
| No. of past-year hospitalization | 7.18 | 2.324 | .010 |
| Sequence | 6.72 | 2.077 | .019 |
| Length of stay | 6.47 | 1.957 | .025 |
| Coronary artery disease | 6.24 | 1.898 | .029 |
| Vascular disease w/c | 6.31 | 1.847 | .032 |
| Admission source | 5.95 | 1.794 | .036 |
| Atrial fibrillation | 6.03 | 1.736 | .041 |
| Priority | 5.77 | 1.705 | .044 |
| Agent Orange | 5.62 | 1.682 | .046 |
| Pneumonia | 5.66 | 1.662 | .048 |
| Sex | 5.24 | 1.656 | .049 |
| Mental disorder | 5.39 | 1.632 | .051 |
| Malignant neoplasm | 5.27 | 1.615 | .053 |

**Table 6** Calibration by readmission risk decile

| Risk decile | Sample size | Predicted PAR | Observed PAR | O/P ratio |
|---|---|---|---|---|
| 1 | 2286 | 201 | 183 | 0.910 |
| 2 | 1112 | 149 | 141 | 0.946 |
| 3 | 893 | 106 | 118 | 1.113 |
| 4 | 481 | 94 | 102 | 1.085 |
| 5 | 343 | 79 | 83 | 1.051 |
| 6 | 215 | 77 | 74 | 0.961 |
| 7 | 138 | 48 | 45 | 0.938 |
| 8 | 82 | 31 | 28 | 0.903 |
| 9 | 29 | 17 | 15 | 0.882 |
| 10 | 16 | 7 | 6 | 0.857 |

Table 7 Performance comparisons of our model over the selected methods

| Method | Predictive accuracy measure | | | | | | | |
|---|---|---|---|---|---|---|---|---|
| | Sensitivity | Specificity | PPV | NPV | F-score | MCC | MSE | AUROC |
| Our proposal | 91.95% | 97.65% | 86.61% | 98.65% | .892 | .874 | .032 | .836 |
| Random Forest | 88.43% | 97.35% | 84.70% | 98.07% | .865 | .843 | .039 | .802 |
| SVM | 86.16% | 97.52% | 85.20% | 97.70% | .857 | .833 | .041 | .775 |
| Logistic Regression | 83.40% | 97.21% | 83.19% | 97.25% | .833 | .805 | .048 | .721 |
| Neural Network | 82.39% | 97.06% | 82.28% | 97.08% | .823 | .794 | .051 | .704 |

---

Algorithm 1. Phase-type Survival Forest

I. For $b = 1$ to $B$:

   a) Take a bootstrap sample (i.e., a random sample chosen with replacement) of size $S$ at the *subject* level (patient) from the training data. Assuming $n_j$ records per patient $j$, we have $N = n_1 + \cdots + n_S$.

   b) Grow an *unpruned* tree $T_b$ on each bootstrap by repeating the following steps, until no improvement is made in $IG_\ell$.

   i. Select $v'$ variables at random from the whole $v$ variables. Normally $v'$ should be much less than $v$, such as $\sqrt{v}$ or even 1.

   ii. Following the splitting criterion introduced, pick the best variable among the $v'$, and split the node into two child nodes.

II. Output the ensemble of trees $\{T_b\}_1^B$.

To make a prediction for a new patient $x$:

Let $\hat{C}_b(x_{(i)})$ be the class prediction of the $b$-th tree for replicate $i$ of the patient. Then we have $\hat{C}(x) =$ majority vote $\{\hat{C}_b(x_{(i)})\}_1^B$.